\documentclass[%
superscriptaddress,
preprint,
 amsmath,amssymb,
prb,
]{revtex4-1}

\usepackage{comment}
\usepackage[dvipdfmx]{graphicx}
\usepackage{dcolumn}
\usepackage{bm}
\usepackage{hyperref}
\begin{document}

\title{Relationships between Superconductivity and Nematicity in FeSe$_{1-x}$Te$_x$ ($x=0-0.5$) Films Studied by Complex Conductivity Measurements}


\author{Hodaka Kurokawa}
\address{Department of Basic Science, The University of Tokyo, 3-8-1 Komaba, Meguro-ku, Tokyo, Japan}
\email[E-mail: ]{kurokawa00128@gmail.com}

\author{Sota Nakamura}
\address{Department of Basic Science, The University of Tokyo, 3-8-1 Komaba, Meguro-ku, Tokyo, Japan}

\author{Jiahui Zhao}
\address{Department of Basic Science, The University of Tokyo, 3-8-1 Komaba, Meguro-ku, Tokyo, Japan}

\author{Naoki Shikama}
\address{Department of Basic Science, The University of Tokyo, 3-8-1 Komaba, Meguro-ku, Tokyo, Japan}

\author{Yuki Sakishita}
\address{Department of Basic Science, The University of Tokyo, 3-8-1 Komaba, Meguro-ku, Tokyo, Japan}

\author{Yue Sun}
\address{Department of Physics, Aoyama Gakuin University, 5-10-1 Fuchinobe,  Chuou-ku, Sagamihara,
	Japan. }

\author{Fuyuki Nabeshima}
\address{Department of Basic Science, The University of Tokyo, 3-8-1 Komaba, Meguro-ku, Tokyo, Japan}

\author{Yoshinori Imai}
\address{Department of Physics, Graduate School of Science, Tohoku University, 6-3 Aramaki-Aoba, Aoba-ku, Sendai, Miyagi, Japan }

\author{Haruhisa Kitano}
\address{Department of Physics, Aoyama Gakuin University, 5-10-1 Fuchinobe,  Chuou-ku, Sagamihara,
	Japan. }

\author{Atsutaka Maeda}
\address{Department of Basic Science, The University of Tokyo, 3-8-1 Komaba, Meguro-ku, Tokyo, Japan}


\begin{abstract}
	We measured the complex conductivity, $\sigma$, of FeSe$_{1-x}$Te$_x$ ($x=0-0.5$) films in the superconducting state which show a drastic increase of the superconducting transition temperature, $T_\textrm{c}$, when the nematic order disappears. Since the magnetic penetration depth, $\lambda$  $(>$  400 nm), of Fe(Se,Te) is longer than the typical thickness of the film ($\sim$100 nm), we combined the coplanar waveguide resonator  and cavity perturbation techniques to evaluate both the real and imaginary parts of $\sigma$. Films with a nematic order showed a qualitatively different temperature dependence in penetration depth and quasiparticle scattering time when compared with those without nematic order, suggesting that nematic order influences the superconducting gap structure. Conversely, the proportionality between  superfluid density, $n_\textrm{s}$ ($\propto\lambda^{-2}$),  and $T_\textrm{c}$ was observed irrespective of the presence or absence of  nematic order. This result indicates that the amount of superfluid has a stronger impact on the $T_\textrm{c}$ of Fe(Se,Te) than the presence or absence of nematic order. Combining these results with band dispersions calculated using density functional theory, we propose that the change of the Fermi surface associated with nematicity is the primary factor influencing the change of $T_\textrm{c}$ and the superconducting gap structure in Fe(Se,Te).
\end{abstract}

\maketitle

\section{Introduction}

Iron chalcogenide superconductor, FeSe, has been intensively studied by virtue of its various intriguing properties:\cite{Liu2015,Bohmer2018,Kreisel2020a} the potential for high-transition-temperature superconductivity, the absence of magnetic order under ambient pressures, and their ability to exhibit exotic electronic states as a result of their extremely small Fermi surface. The superconducting transition temperature, $T_\textrm{c}$, can be enhanced above 40 K from 9 K by intercalation,\cite{Hsu2008,Burrard-Lucas2013} carrier doping using an electron double layer transistor,\cite{Shiogai2016, Shikama2020} and synthesis of a monolayer film.\cite{Wang2012, Huang2017} The nematic phase without magnetic order in FeSe is ideal for studying the origin of nematicity  and the relationship between nematicity and superconductivity.\cite{Fernandes2014, Glasbrenner2015} Furthermore, the small Fermi surface  ($\epsilon_\textrm{F}< 10$ meV)  can easily be tuned by hydrostatic pressure,\cite{Margadonna2009} chemical pressure via isovalent substitution,\cite{Fang2008,Mizuguchi2009,Imai2015} and the in-plane lattice strain.\cite{Nabeshima2018b} 
Since changes in the Fermi surface influence the superconducting, nematic, and magnetic phases, various techniques have been applied to investigate the electronic phase diagram and exotic superconductivity of FeSe.

Among the above-mentioned techniques to control the electronic state, chemical isovalent substitution is advantageous since experiments can be performed under ambient pressures. The S-substitution shrinks the lattice of FeSe, resulting in positive chemical pressure. With increasing S content, the nematic transition temperature, $T_\textrm{n}$, decreases, and $T_\textrm{c}$ slightly increases and decreases.\cite{Watson2015} Although no significant changes in $T_\textrm{c}$ occur when the nematic order disappears, some abrupt changes in the superconducting gap have been observed in a measurement of thermal properties and the scanning tunneling microscopy/spectroscopy.\cite{Sato2018,Hanaguri2018} Hence, the nematic order or its fluctuation may exert some influence on the superconducting state. 
Conversely, few systematic investigations of Te-substituted FeSe, which are subject to negative chemical pressures, have been conducted relative to those concerning Fe(Se,S) since the systematic synthesis of bulk Fe(Se,Te) had, until recently, been hindered by the phase separation region.\cite{Terao2019}  Since  the superconducting gap structure of FeSe is distinctly different from that of  FeSe$_{1-x}$Te$_x$ ($x>0.5$),\cite{Hanaguri2008,Kasahara2014,Sprau2017} it is necessary to understand how the superconducting gap evolves with increasing Te content.

Before the systematic synthesis of bulk Fe(Se,Te), we succeeded in growing  single-crystalline thin films of FeSe$_{1-x}$Te$_x$ in the whole composition ($x=0-0.9$) using a pulsed laser deposition technique.\cite{Imai2015,Imai2017} Although the $T_\textrm{n}$ of the Fe(Se,Te) films decreased after Te substitution, $T_\textrm{c}$ was largely enhanced after the disappearance of nematic order.\cite{Imai2015,Bellingeri2010,Sylva2018} This enhancement of $T_\textrm{c}$ is contrary to the Fe(Se,S) films and bulk Fe(Se,Te),\cite{Nabeshima2018a,Terao2019} indicating that the effect of nematicity on $T_\textrm{c}$ is complicated in these materials. Although a positive correlation between $T_\textrm{c}$ and the carrier density or the density of states (DOS) has been reported through magneto-transport and the angle-resolved photoemission spectroscopy (ARPES) in the normal state,\cite{Nabeshima2020,Nakayama2021} the superconducting properties of these films and their relation to nematicity are yet to be fully understood. To elucidate the effects of Te substitution and nematicity on superconductivity, we investigated both the response of the superfluid and dynamics of quasiparticles below $T_\textrm{c}$ in Fe(Se,Te) films. 


In this paper, we report systematic measurements of the complex conductivity, $\sigma$, of FeSe$_{1-x}$Te$_x$ ($x=0-0.5$) films below $T_\textrm{c}$. Since the magnetic penetration depth, $\lambda$, is several times as long as the typical thickness of the film ($\sim100$ nm), measurement techniques applicable to  bulk crystals cannot be used. Hence, to evaluate both the real and imaginary parts of $\sigma$, we combined the coplanar waveguide resonator and cavity perturbation techniques. The quasiparticle scattering time, $\tau$, was calculated from the real part of $\sigma$ and was found to increase at low temperatures, as observed in bulk FeSe and FeSe$_{0.4}$Te$_{0.6}$.\cite{Takahashi2011,Okada2017} Moreover, the $\lambda$ and $1/\tau$ of films with nematic order showed a quantitatively distinct behavior from films without nematic order, suggesting that changes in the superconducting gap structure are associated with nematic order. Conversely, the proportionality between  superfluid density, $n_\textrm{s}(\propto\lambda^{-2})$, and $T_\textrm{c}$ was observed irrespective of the presence or absence of nematic order. 
Additionally, using density functional theory (DFT) calculations, we confirmed a change of Fermi surface associated with nematic order, which is considered to influence the superconducting gap structure. Moreover, DOS decreased in the nematic phase, which may have caused a decrease in superfluid density. From these results, we suggest that the change of the band structure in the nematic phase primarily influences the superconducting gap structure and $T_\textrm{c}$ rather than the nematic fluctuation developing near the nematic quantum critical point.\cite{Mukasa2021}


\section{Experiments}

\subsection{Sample}
All films were grown on CaF$_2$ substrates  ($\sim5\times5\times0.5$ mm$^3$) via the pulsed laser deposition method using a KrF laser. The details of film growth have been  described elsewhere.\cite{Imai2010,Imai2010a} The thicknesses of grown films were measured using a stylus profiler. DC electrical resistivity was measured using a standard four-probe method equipped with a physical property measurement system (Quantum Design, PPMS). 

\begin{figure}
	\begin{center}
		\includegraphics[width=90 mm]{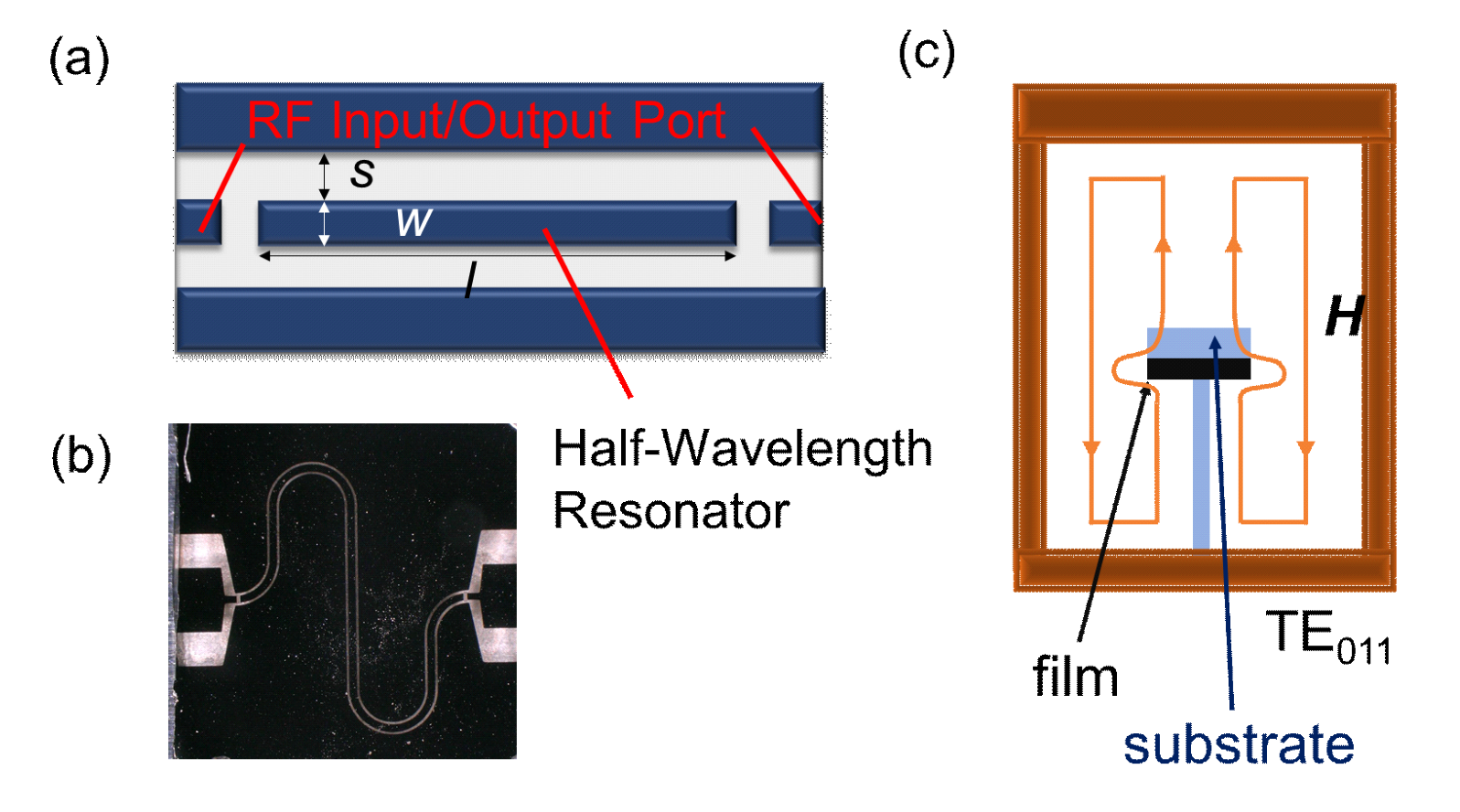}
	\end{center}
	\caption{(a) Schematic of the coplanar resonator. (b) The fabricated coplanar resonator (FeSe$_{0.8}$Te$_{0.2}$). (c) Schematic of the cavity resonator.}
	\label{fig:coplanar}
\end{figure}

\subsection{Measurements and calculations}
To measure $\lambda$ of FeSe$_{1-x}$Te$_x$ films, we fabricated the $\sim5\times5 $ mm$^2$ film into the coplanar waveguide resonator (Fig. \ref{fig:coplanar} [a]) by Ar ion milling and focused ion beam (FIB). Ar ion milling was used to fabricate the whole structure, after which the 50 $\mu$m gap between the resonator and the microwave input/output port was etched using FIB.  The width of the resonator, $w$, the gap between the resonator and the ground, $s$, and the length of the resonator, $l$, were designed to be 120 $\mu$m, 30 $\mu$m, and 6.2-9.9 mm, respectively. The resonance frequency of the coplanar resonator was 6-10 GHz.
Figure \ref{fig:coplanar} (b) shows the fabricated resonator on the FeSe$_{0.8}$Te$_{0.2}$ film. The resonator was mounted onto a printed circuit board, which was connected to the resonator by Al wirebonding. This was cooled down to 2 K using PPMS. Transmitted power was measured using a network analyzer (Keysight, N5222A).

$\lambda$ was calculated from the resonance frequency, $f_\textrm{c}$, as follows. 
For the half-wavelength coplanar resonator: 
\begin{equation}\label{fc}
f_\textrm{c}=\frac{1}{2l\sqrt{LC}},
\end{equation}
where $L$ is the inductance per unit length and $C$ is the capacitance per unit length. Using an electromagnetic simulation software (WIPL-D), we confirmed that the coupling between the resonator and the input port had negligible effects on $f_\textrm{c}$. For a superconductor: 
\begin{equation}\label{L}
L=L_\textrm{m}+L_\textrm{k},
\end{equation}
where $L_\textrm{m}$ is the magnetic inductance and $L_\textrm{k}$ is the kinetic inductance corresponding to the response of the superfluid \cite{Watanabe1994}. $L_\textrm{k}$ is a quadratic function of $\lambda$ as follows:
\begin{equation}\label{Lk}
L_\textrm{k}=\frac{\mu_0g(s,w,d)}{dw}\lambda^2,
\end{equation}
where $\mu_0$ is vacuum permeability, $g(s,w,d)$ is a geometrical factor,  and $d$ is the thickness of the film \cite{Watanabe1994,Clem2013}. From eq. (\ref{fc})-(\ref{Lk}), $\lambda$ is expressed as follows:
\begin{equation}\label{lambda}
\lambda = \sqrt{\frac{dw}{g\mu_0}\left( \frac{1}{4l^2Cf_\textrm{c}^2}-L_\textrm{m}\right) }.
\end{equation}
All parameters on the right-hand side of eq. (\ref{lambda}) can be determined from the shape of the resonator ($s$, $w$, $d$) and measurements of $f_\textrm{c}$ and $C$. The length was measured using an optical microscope (Keyence, VHS-6000), and the thickness was measured using a stylus profiler as mentioned above. The typical standard deviations of $s$ and $w$ and $d$ were about 1-2 um, 1-2 um and ~1 nm, respectively, leading to the standard deviation of $\lambda$ of ~50nm, which was shown as an error bar. Also, $C$ was measured using an impedance analyzer (Hewlett-Packard, 4192A) in the frequency range, 10-1000 kHz. The measured values of $C$ were in good agreement with the calculated values of $C$ assuming that the relative permittivity of CaF$_2$ is 6.5. \cite{Jacob2003}


The dynamics of quasiparticles in the Fe(Se,Te) films were measured using the cavity perturbation technique. For a thin film ($d<\lambda$), the cavity perturbation formula for the analysis of a bulk crystal cannot be applied. In such cases, the measured quantity is the effective impedance, $Z_\textrm{eff} (Z_\textrm{s}, d)$, where $Z_\textrm{s}$ is surface impedance. Formulae of $Z_\textrm{eff}$ corresponding to various situations have been derived, and are found to depend on configurations of  both the electromagnetic field and the sample.\cite{Klein1990,Drabeck1990, Peligrad1998, Barannik2014}

A flake of FeSe$_{1-x}$Te$_x$ film with substrate was cut from the coplanar resonator after the measurement of $\lambda$. This flake ($\sim$ 0.5 $\times$ 0.5 mm$^2$) was mounted onto a sapphire rod at the center of the cavity resonator (Fig. \ref{fig:coplanar} [c]). The TE$_{011}$ mode (44 GHz) of the resonator was used with a configuration in which the magnetic field of the TE$_{011}$ mode was parallel to the film well below $T_\textrm{c}$ such that $Z_\textrm{eff}$ can be expressed as follows:
\begin{equation}\label{Zeff}
Z_\textrm{eff}=-\frac{i}{2}Z_\textrm{s}\textrm{cot}\left(\frac{\omega\mu_0d}{2Z_\textrm{s}}\right),
\end{equation}
where $\omega$ is angular frequency \cite{Barannik2014}. Experimentally, the effective surface resistance, $R_\textrm{eff}$, is determined by the following equation:
\begin{equation}\label{Reff}
R_\textrm{eff}=G\left(\frac{1}{2Q_\textrm{sample}}-\frac{1}{2Q_\textrm{blank}}  \right),
\end{equation}
where $G$ is the geometric factor, $Q_\textrm{sample}$ is the quality factor of the cavity containing the sample, $Q_\textrm{blank}$ is the quality factor of the cavity without the sample. Here, we have confirmed that the effect of the CaF$_2$ substrate was negligible by measurement of the substrate alone. 
Also, the effective surface reactance, $X_\textrm{eff}$ is as follows:
\begin{multline}\label{Xeff}
X_\textrm{eff}(T)=G\left(\frac{f_\textrm{c,sample}(T_0)-f_\textrm{c,sample}(T)}{f_\textrm{c,sample}(T_0)}-\right.\\
\left.\frac{f_\textrm{c,blank}(T_0)-f_\textrm{c,blank}(T)}{f_\textrm{c,blank}(T_0)}          \right) + X_\textrm{eff}(T_0),
\end{multline}
where $f_\textrm{c,sample}$ is the resonance frequency with the sample, $f_\textrm{c,blank}$ is the resonance frequency without the sample, and $T_0$ is the minimum temperature during the measurement, typically 2.1 K. %

To obtain $Z_\textrm{s}$ by solving eq. (\ref{Zeff})-(\ref{Xeff}), we determined $G$ and $X_\textrm{eff}(T_0)$ as follows. At low temperatures where $\sigma_1<<\sigma_2$,
\begin{equation}\label{Xeff_lowT}
X_\textrm{eff}(T)=\frac{1}{2}\mu_0\omega\lambda\textrm{coth}\left(\frac{d}{2\lambda}\right)
\end{equation}
from eq. (\ref{Zeff}) and $X_\textrm{s}\approx\mu_0\omega\lambda$. Thus, $X_\textrm{eff}(T)$ can be calculated by substituting the value of $\lambda(T)$ measured by the coplanar resonator into eq. (\ref{Xeff_lowT}). Here, $X_\textrm{eff}(T_0)$ was obtained using eq. (\ref{Xeff_lowT}) and $\lambda(T_0)$ measured by the coplanar resonator. Conversely, $G$ was determined by curve fitting assuming that $X_\textrm{eff}(T)$ calculated using eq.(\ref{Xeff_lowT}) and $\lambda(T)$ measured by the coplanar resonator and $X_\textrm{eff}(T)$ obtained from eq.(\ref{Xeff}) and the measurement of the cavity resonator is equal in the temperature range,  $0.2-0.5 T_\textrm{c}$, where the approximation, $X_\textrm{s}\approx\mu_0\omega\lambda$, hold. After determining $G$ and $X_\textrm{eff}(T_0)$,  we numerically solved eq. (\ref{Zeff}) and obtained $Z_\textrm{s}$. It should be noted that eq. (\ref{Zeff}) is not applicable near $T_\textrm{c}$ due to the drastic change of the electromagnetic field distribution around the film.\cite{Barannik2014} Here, we determined the upper temperature limit for an applicable range of eq. (\ref{Zeff}) to be below 0.75 $T_\textrm{c}$ from a measurement of a conventional superconductor, NbN film.

Besides the measurement of $\sigma$, we performed DFT calculation using FPLO-18. The exchange functional was a generalized gradient approximation (GGA+U). The $k$-mesh was $12\times12\times6$. For the calculation of the nematic phase, we applied the technique proposed in Ref. \cite{Long2020}, using lattice constants for FeSe as follows: $a=3.76976$ \AA, $c=5.52122$ \AA  and $z_\textrm{Se}=0.2688$.\cite{Glasbrenner2015} 

\section{Results and Discussion}
\subsection{Measurements of complex conductivity}
\begin{figure}
	\begin{center}
		\includegraphics[width=80 mm]{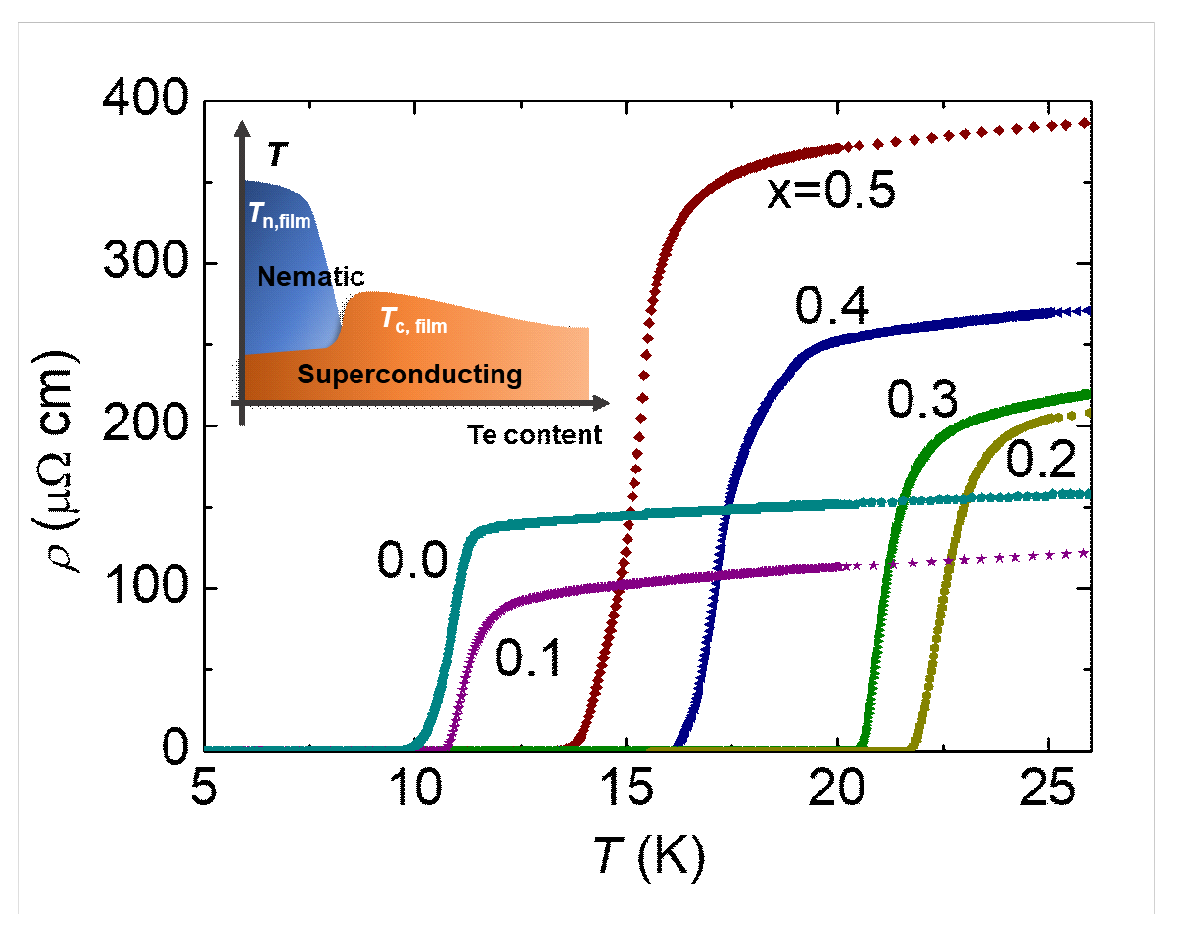}
	\end{center}
	\caption{Temperature dependence of dc resistivity of the FeSe$_{1-x}$Te$_x$ ($x=0-0.5$) films. The inset shows a schematic of the phase diagram of FeSe$_{1-x}$Te$_x$ films.\cite{Imai2015} }
	\label{fig:RT}
\end{figure}
Figure \ref{fig:RT} shows the temperature dependence of DC resistivity in FeSe$_{1-x}$Te$_x$ ($x=0-0.5$) films.  $T_\textrm{c}$ increased from $x=0$ to $x=0.2$, consistent with previous reports.\cite{Imai2015} Subsequently, $T_\textrm{c}$ gradually decreased with increasing Te content, and the resistivity of $T_\textrm{c,onset}$ increased from $\sim$ 100 $\mu\Omega$ cm at $x=0$ to $\sim$ 400 $\mu\Omega$ cm at $x=0.5$, which was again a typical value for these films.\cite{Imai2017}
\begin{figure}
	\begin{center}
		\includegraphics[width=86 mm]{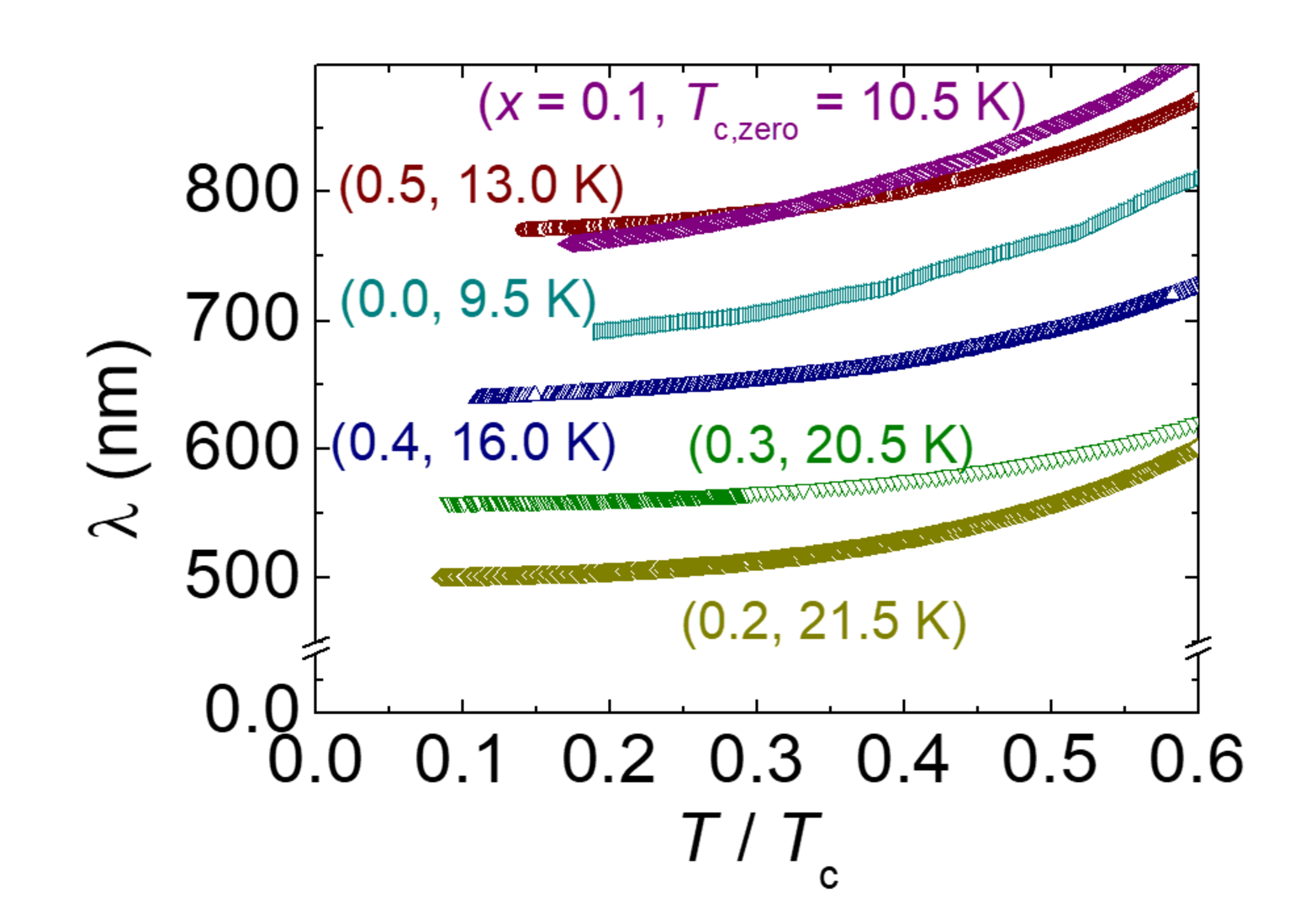}
	\end{center}
	\caption{$\lambda$ of FeSe$_{1-x}$Te$_x$ ($x=0.0-0.5$) films as a function of reduced temperature. $T_\textrm{c, zero}$ of each film is also shown.}
	\label{fig:lambda-all}
\end{figure}

\begin{figure}
	\begin{center}
		\includegraphics[width=86 mm]{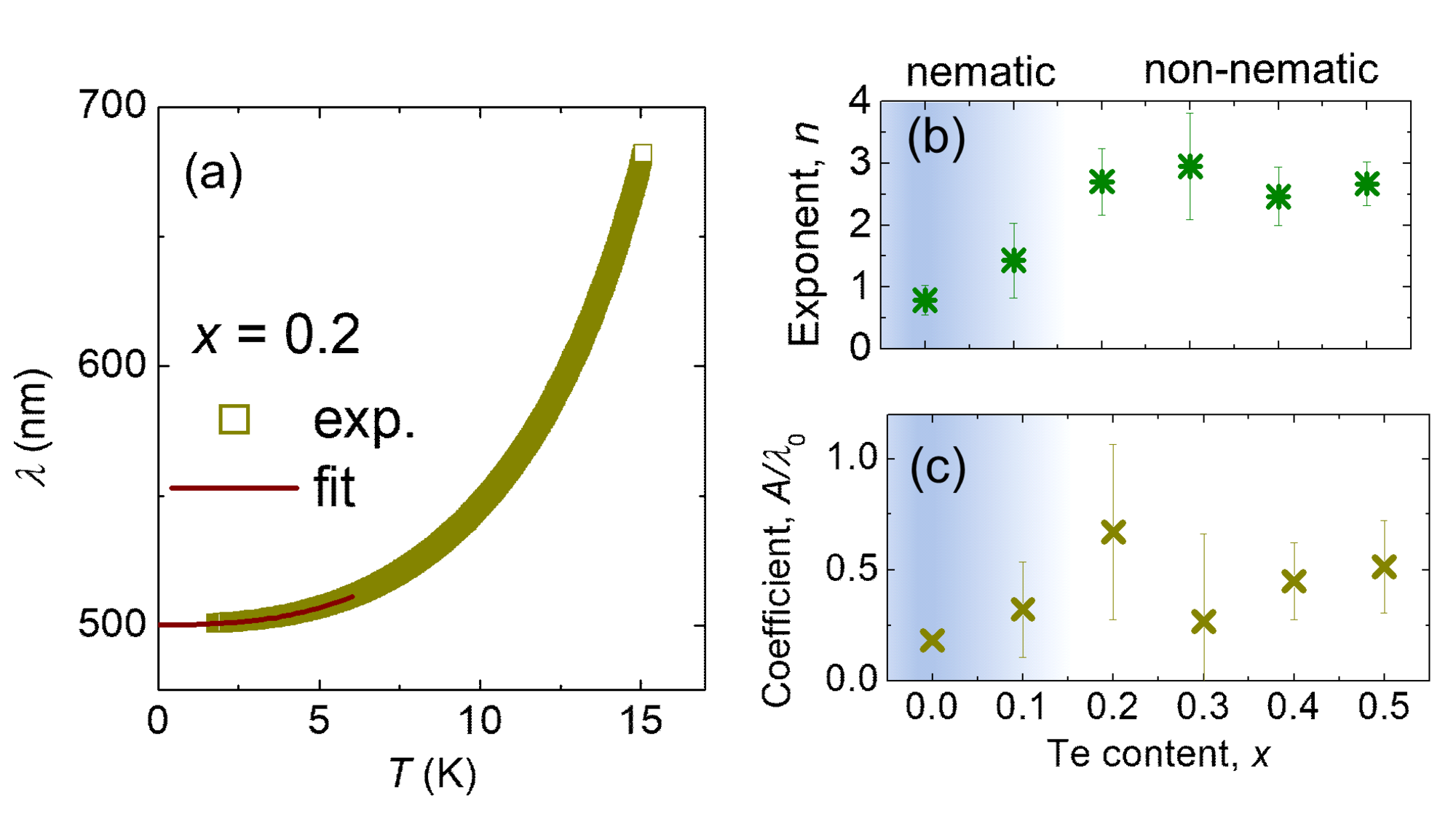}
	\end{center}
	\caption{(a) Temperature dependence of $\lambda$ of FeSe$_{0.8}$Te$_{0.2}$ film. The red line corresponds a fitted curve with the equation, $\lambda(T)=\lambda_0+A(T/T_\textrm{c})^n$. (b) $n$ and (c) $A/\lambda_0$ as a function of Te content.}
	\label{fig:lambda-fit}
\end{figure}

\begin{figure}
	\begin{center}
		\includegraphics[width=86 mm]{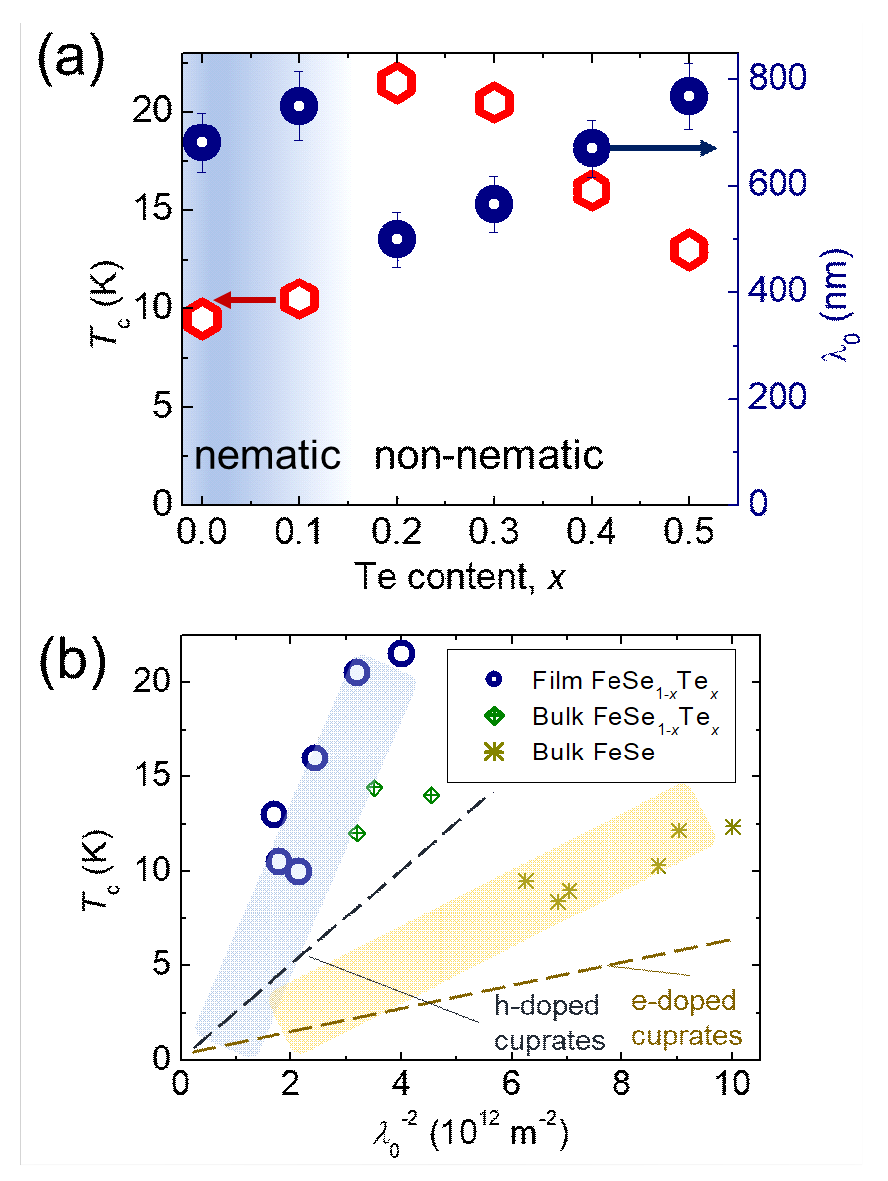}
	\end{center}
	\caption{(a) $T_\textrm{c}$ and $\lambda$ at 0 K in FeSe$_{1-x}$Te$_x$ ($x=0-0.5$) films. (b) $T_\textrm{c}$ as a function of $\lambda_0^{-2}$ in FeSe$_{1-x}$Te$_x$ ($x=0-0.5$) films. Results of bulk FeSe and bulk FeSe$_{1-x}$Te$_x$ ($x>0.5$) are also shown.\cite{Khasanov2010,Kasahara2014, Okada2017,Takahashi2011,Kim2010,Biswas2010} The blue line and yellow dashed lines correspond to the data of hole-doped cuprates and electron-doped cuprates, respectively.\cite{Luetkens2008a}}
	\label{fig:lambda}
\end{figure}
Using the temperature dependence of $f_\textrm{c}(T)$ of the coplanar resonator, we calculated $\lambda (T)$ using eq. (\ref{lambda}). Fig. \ref{fig:lambda-all} shows $\lambda$ of FeSe$_{1-x}$Te$_x$ ($x=0.0-0.5$) films as a function of reduced temperature. The obtained $\lambda(T)$ was extrapolated to 0 K assuming that $\lambda(T)=\lambda_0+A(T/T_\textrm{c})^n$, where $\lambda_0$ is the penetration depth at 0 K, and $A$ and $n$ are constants. Here, we performed curve fitting in the temperature range from 2 K to 0.3 $T_\textrm{c}$. \cite{Prozorov2011} Fig. \ref{fig:lambda-fit} (a) shows the fitting result of FeSe$_{0.8}$Te$_{0.2}$ as an representative.

Figure \ref{fig:lambda-fit} (b) shows $n$ of FeSe$_{1-x}$Te$_x$ films. For films in nematic phase ($x$=0, 0.1), $n$ was almost 1, whereas $n$ was 2$-$3 in non-nematic phase. The $T$-linear behavior indicates that the superconducting gap has nodes or gap minima in nematic phase. \cite{Kasahara2014,Okada2017}. Conversely, $n>2$ in non-nematic phase probably denote the existence of nodeless gaps which are subjected to pair-breaking effects.\cite{Prozorov2011}
On the other hand, no systematic changes in the dimensionless coefficient, $A/\lambda_0$, were observed as shown in Fig.\ref{fig:lambda-fit} (c). For cases such as $d$-wave superconductor with line nodes or gapless superconductor, $A/\lambda_0$ can be related to the superconducting gap or some other material properties.\cite{Prozorov2011} However, in the case of Fe(Se,Te) films, superconducting gap structure changes with increasing Te content. Thus, though we show $A/\lambda_0$ for the clarity of fitting results, interpreting $A/\lambda_0$ for all samples in terms of any universal standpoint is difficult, rather does not make sense.

Figure \ref{fig:lambda} (a) shows Te content \textit{versus} $T_\textrm{c,zero}$ and $\lambda_0$. The negative correlation between $T_\textrm{c}$ and $\lambda_0$ seems to exist irrespective of the presence or absence of nematic order. Subsequently, we plotted $T_\textrm{c}$ as a function of $\lambda_0^{-2}$ (Fig. \ref{fig:lambda} [b]), which is the so-called Uemura plot. $T_\textrm{c}$ exhibited an obvious positive correlation with $\lambda_0^{-2}$, corresponding to $n_\textrm{s}$. The observed proportionality between $T_\textrm{c}$ and $n_\textrm{s}$ is consistent with the correlation between $T_\textrm{c}$ and the carrier density of the Fe (Se,Te) and Fe(Se,S) films in their normal state.\cite{Nabeshima2020} These results indicate that either  $n_\textrm{s}$ or carrier density plays a crucial role in determining the $T_\textrm{c}$ of Fe(Se,Te) films irrespective of the presence or absence of nematic order or its fluctuation. Of note, this kind of correlation between $T_\textrm{c}$ and $\lambda_0^{-2}$ was widely observed in other iron based superconductors, suggesting that superconductivity is induced by electronic correlation in these materials.\cite{Uemura1989,Bendele2010,Rodiere2012}

We remark that an abrupt change of $\lambda_0$ with increasing disorder was reported in Ba(Fe,Rh)$_2$As$_2$, indicating that the change of superconducting gap structure from $s\pm$ to $s++$.\cite{Ghigo2018}  Also in Fe(Se,Te) films, when superconducting gap structure changed from nodal (nematic phase) to nodeless (non-nematic phase), $\lambda_0$ changed abruptly. Thus, it can be said not only $\lambda(T)$ but also $\lambda_0$ is sensitive to the change of the superconducting gap structure.

Compared with bulk samples, whereas the trend between $T_\textrm{c}$ and $\lambda_0^{-2}$ in the films was similar to that of bulk FeSe$_{1-x}$Te$_x$ ($x>0.5$),\cite{Takahashi2011, Kim2010} a discrepancy with bulk FeSe was observed.\cite{Kasahara2014, Okada2017} Namely, the slope of the data of FeSe$_{1-x}$Te$_x$ ($x=0-0.5$) films and bulk FeSe$_{1-x}$Te$_x$ ($x>0.5$) was larger than that of bulk FeSe. Such differences in the Uemura plot have already been reported in cuprate superconductors, in which the data of hole-doped cuprates show the larger slopes than that of electron-doped cuprates (dotted lines in Fig. \ref{fig:lambda} [b]). Interestingly, whereas $n_\textrm{h}= 1.1-1.4n_\textrm{e}$ in the bulk FeSe,\cite{Ovchenkov2018} $n_\textrm{h}=1.0-2.8n_\textrm{e}$ in the FeSe$_{1-x}$Te$_x$ films which showed the steeper slope,\cite{Sawada2016,Yoshikawa2019,Nabeshima2020} where $n_\textrm{h}$ is the hole density and $n_\textrm{e}$ is the electron density. The carrier density of the bulk as-grown FeSe$_{1-x}$Te$_x$ ($x>0.5$) is also estimated to be $n_\textrm{h}>n_\textrm{e}$ from measurements of the Hall coefficient.\cite{Liu2009, Sun2014} This correspondence between Fe(Se,Te) and the cuprates suggests the possibility that hole-doping increases the slope of the Uemura plot, even in multi-band superconductors.





\begin{figure*}
	\begin{center}
		\includegraphics[width=172 mm]{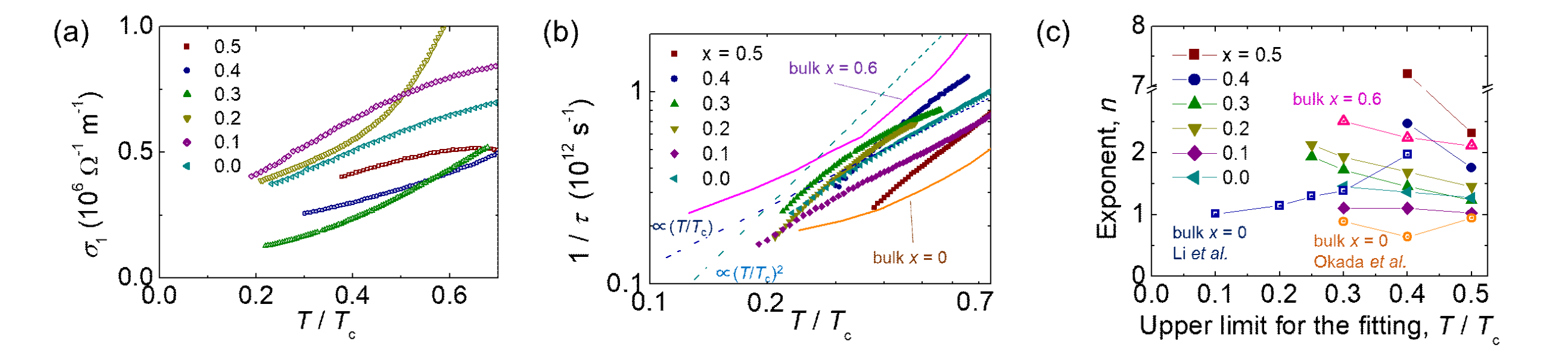}
	\end{center}
	\caption{(a) Temperature dependence of $\sigma_1$ and (b) the inverse of $\tau$ in the  FeSe$_{1-x}$Te$_x$ ($x=0-0.5$) films. The result of bulk crystals are also shown \cite{Takahashi2011, Okada2017}. (c) The exponent $n$ in the equation, $1/\tau=aT^n+b$, determined from curve fitting. The maximum temperature for the fitting was varied from 0.25$T/T_\textrm{c}$ to 0.5$T/T_\textrm{c}$. Orange circles and blue squares correspond to $1/\tau$ of the bulk FeSe \cite{Okada2017,Li2016}. Pink triangles denote $1/\tau$ of the bulk FeSe$_{0.4}$Te$_{0.6}$ \cite{Takahashi2011}. }
	\label{fig:tau}
\end{figure*}
Next, we evaluated the results of measurements of the dynamics of quasiparticles using the cavity perturbation technique. 
From the measurement of the $Q^{-1}(T)$ and $f_\textrm{c}(T)$ in each films placed in the cavity resonator, $R_\textrm{s}$ and $X_\textrm{s}$ were calculated using eq. (\ref{Zeff}) below $0.75\ T_\textrm{c}$.
The real part of the complex conductivity, $\sigma_1$, was calculated using $\sigma_1=2\omega\mu_0R_\textrm{s}X_\textrm{s}/(R_\textrm{s}^2+X_\textrm{s}^2)^2$. When calculating $\sigma_1$, we subtracted residual surface resistance from $R_\textrm{s}$, which was estimated from the linear extrapolation of $R_\textrm{s}$ to 0 K.  Figure \ref{fig:tau} (a) shows the value of $\sigma_1$ corresponding to each film. With decreasing temperature, $\sigma_1$ decreased in all tested films. The decrease of $\sigma_1$ at low temperatures is consistent with the measurement of bulk FeSe and  FeSe$_{0.4}$Te$_{0.6}$.\cite{Li2016,Takahashi2011}

Assuming the two-fluid model and Drude-like single-carrier normal fluid, the quasiparticle scattering time, $\tau$, can be expressed as follows:
\begin{equation}\label{tau}
\omega\tau = \frac{\tilde{\sigma_1}}{1-\tilde{\sigma_2}},
\end{equation}
where $\tilde{\sigma}= \tilde{\sigma_1}+i\tilde{\sigma_2}=\mu_0\omega\lambda_0^2(\sigma_1+i\sigma_2)$, which gives the dimensionless conductivity.\cite{Takahashi2011}
Since temperature dependence of $\tau$ did not strongly depend on values of residual surface resistance, we used $\tau$ to compare intrinsic properties of Fe(Se,Te) films instead of $\sigma_1$. Here, we should be careful this single-carrier treatment because Fe(Se,Te) is a multi-band superconductor. Since FeSe has highly anisotropic gaps in both hole and electron pockets,\cite{Sprau2017} whereas FeSe$_{1-x}$Te$_{x}$ ($x>0.5$) has nodeless gaps in both pockets ,\cite{Hanaguri2008} the corresponding values $\tau$ of the electron pockets are expected to show similar temperature dependence to that of the hole pocket in FeSe$_{1-x}$Te$_{x}$ ($x=0-0.9$). Hence, in eq. (\ref{tau}), we assumed that the temperature dependence of $\tau$ in both pockets could be captured using a single $\tau$ as a first approximation. 

In all films, $1/\tau$ was observed to decrease at low temperatures (Fig. \ref{fig:tau} [b]), indicating rapid suppression of the inelastic scattering of the electron, which was already established in bulk FeSe and FeSe$_{0.4}$Te$_{0.6}$.\cite{Okada2017,Takahashi2011} In Fe(Se,Te) films, the values of $1/\tau$ at low temperatures were $\sim0.2$, which was similar value to bulk 122 compounds, $1/\tau=0.1\sim0.3$  \cite{Hashimoto2009, Torsello2019}. This coincidence with 122 compounds suggests that our film samples were as clean as 122 materials.  Furthermore, the slope of $1/\tau$ seems to be different among these films as shown in Fig. \ref{fig:tau} (b). To obtain further insights,  we performed curve fitting with $1/\tau=aT^n+b$, where $a$, $b$, and $n$ are positive constants. Figure \ref{fig:tau} (c) shows $n$ of each film as a function of the maximum temperature used for the curve fitting, $T_\textrm{max}^\textrm{fit}$.  The exponent, $n$, showed differing behavior among these films when $T_\textrm{max}^\textrm{fit}$ was decreased. Although $n$ remained constant around 1 in $x=$0, 0.1 films below $T_\textrm{max}^\textrm{fit}=0.5 T_\textrm{c}$, $n$ tended to increase with decreasing $T_\textrm{max}^\textrm{fit}$ in the other films, resulting in $n\geq2$ at low $T_\textrm{max}^\textrm{fit}$. Additionally, in bulk FeSe,\cite{Okada2017,Li2016} which is in the nematic phase same as $x=$0 0.1 films, $n$ was almost equal to 1. In bulk FeSe$_{0.4}$Te$_{0.6}$,\cite{Takahashi2011} which does not show  nematic order, $n$ increased to over 2 with decreasing $T_\textrm{max}^\textrm{fit}$. As was pointed out by Li $et\ al.$,\cite{Li2016} the $T$-linear behavior ($n=1$) in $1/\tau$ may be the consequence of gap structure with line nodes or deep gap minima.\cite{Li2016,Hirschfeld1993,Ozcan2006} Conversely,  the $n>2$ behavior observed for $x\geq0.2$ films could denote the nodeless superconducting gap, since an exponential decrease in $1/\tau$ is expected in nodeless superconductor.\cite{Quinlan1994,Hashimoto2009} Hence, considering the variation in $n$ in samples with different Te contents, we consider that the  superconducting gap structure changes from those with line nodes or deep minima in nematic phase to those that are nodeless outside the nematic phase. The result is consistent with other measurement techniques claiming that bulk FeSe has line nodes or the deep minima, \cite{Sprau2017, Sun2017a} whereas bulk FeSe$_{1-x}$Te$_{x}$ ($x>0.5$) shows nodeless superconducting gaps.\cite{Hanaguri2008}

\subsection{DFT calculations}
\begin{figure}
	\begin{center}
		\includegraphics[width=86 mm]{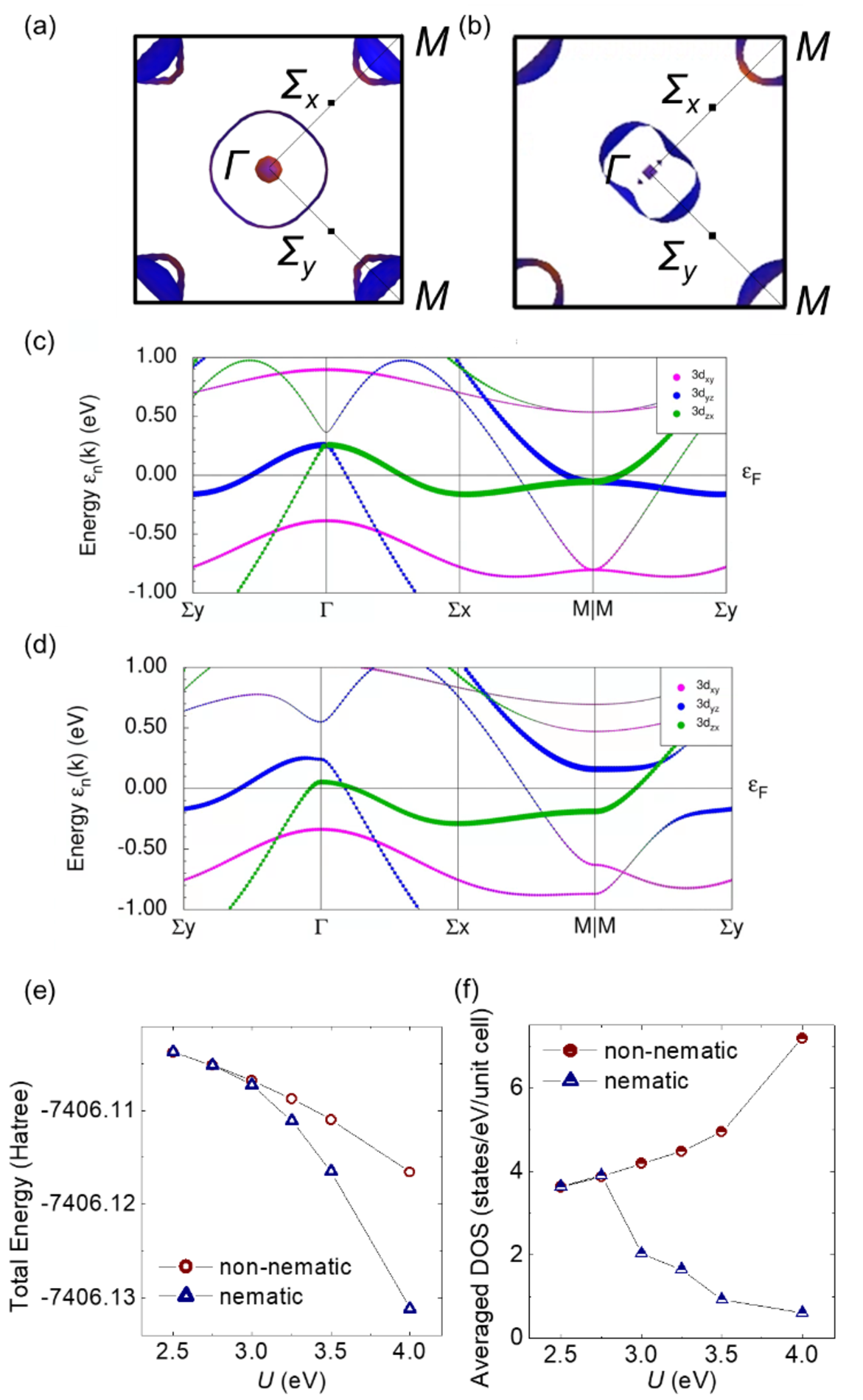}
	\end{center}
	\caption{Fermi surface of FeSe in the (a) non-nematic phase and (b) nematic phase obtained by DFT calculations. Band dispersion of FeSe in the (c) non-nematic phase and (d) nematic phase. (e) Ground state energy of FeSe as a function of $U$ in the non-nematic and nematic phases. (f) Averaged density of states per unit cell of FeSe as a function of $U$ in the non-nematic and nematic phases.
	 }
	\label{fig:DFT}
\end{figure}

To interpret our experimental results with respect to  electronic band structure, 
we calculated the band dispersion of FeSe in both non-nematic and nematic phases. Figure \ref{fig:DFT} (a) and (b) show the Fermi surface of FeSe in the non-nematic and nematic phases calculated using $U = 3$ eV. Figs. \ref{fig:DFT} (c) and (d) show band dispersion along the points depicted in Figs. \ref{fig:DFT} (a) and (b).  In the nematic phase, the Fermi surface shows  two-fold symmetry and the disappearance of one of the electron pockets at the $M$ point (Fig. \ref{fig:DFT} [b]). 
Although the size of Fermi surface was larger than that of experimentally observed one as it is widely accepted,\cite{Liu2015,Coldea2018}
the band dispersion agrees with the previous calculation and qualitatively captures the experimentally observed band structure.\cite{Long2020, Yi2019}

The disappearance of the electron pocket in the nematic phase has been confirmed by several photoemission experiments.\cite{Rhodes2018, Yi2019,Rhodes2020} It has been previously reported that, when considering the absence of one  electron pocket, the strongly anisotropic superconducting gap of FeSe can be reproduced via calculation of the superconducting gap equation.\cite{Rhodes2018} Based on these results, the observed nodal gap in FeSe$_{1-x}$Te$_x$ films with nematic order is probably caused by the disappearance or shrinkage of the electron pocket due to the nematic transition. 

In Fig. \ref{fig:DFT} (e), the ground state energies of the non-nematic and nematic phases are shown. The nematic phase became energetically favorable compared with the non-nematic phase when $U$ increased to values greater than 3 eV.  Furthermore, we compared the averaged DOS near the Fermi surface, which determines the carrier density. Figure \ref{fig:DFT} (f) shows DOS averaged over $\epsilon_\textrm{F}\pm10$ meV in the non-nematic and nematic phases, where  $\epsilon_\textrm{F}$ is the Fermi level. The averaged DOS in the nematic phase is considerably lower than that in the non-nematic phase because of the disappearance of one electron pocket. The difference in averaged DOS between the non-nematic and nematic phases became larger with increasing $U$. The decrease of DOS near the Fermi surface should result in an observed reduction of normal-carrier density in the nematic phase.\cite{Nabeshima2020} Superfluid density is also considered to be reduced in the nematic phase, resulting in a decrease of $T_\textrm{c}$ in the FeSe$_{1-x}$Te$_x$ films in nematic phase. Of note, nematic order disappears at different Te content in Fe(Se,Te) film on CaF$_2$ substrate, LaAlO$_3$ substrate and bulk crystal.\cite{Imai2017,Terao2019} This difference is possibly due to existence of compression strain in Fe(Se,Te) films, which results in the difference in electronic phase diagram of Fe(Se,Te) film and bulk crystal.


\begin{figure}
	\begin{center}
		\includegraphics[width=86 mm]{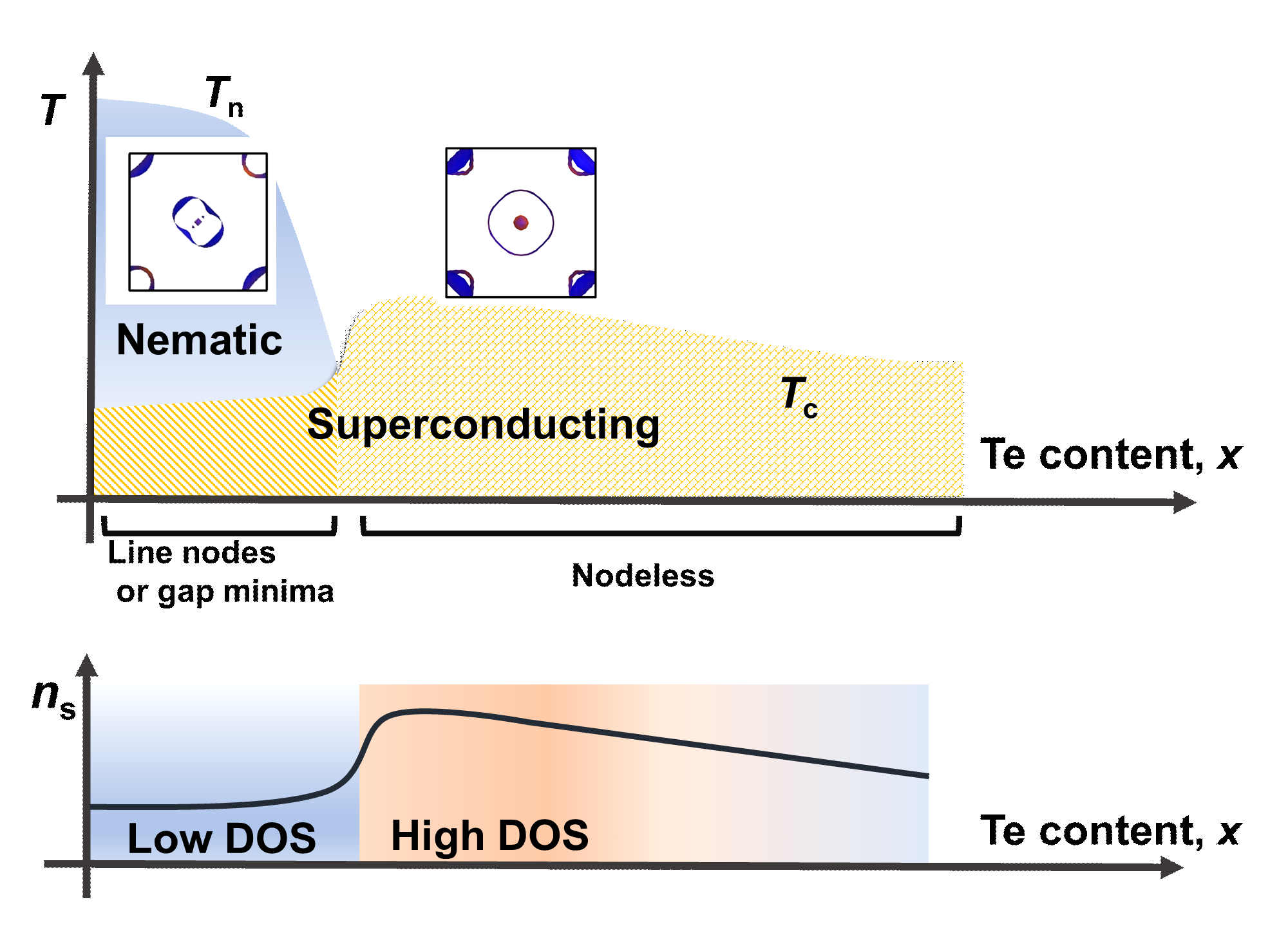}
	\end{center}
	\caption{Schematic phase diagram of FeSe$_{1-x}$Te$_x$ ($x=0-0.5$) films, summarizing the experimental and calculated results.
	}
	\label{fig:summary}
\end{figure}

Figure \ref{fig:summary} summarizes the above results and discussion. The deformation of the Fermi surface in the nematic phase induces strongly anisotropic superconducting gaps with nodes or gap minima. We are not able to determine whether the change in the gap structure takes place exactly at the nematic end point composition or not, since it is rather difficult to change Te content so finely to cover the nematic end point almost continuously.  However, it can be mentioned safely that there is the gross correspondence between the presence/absence of the nematicity and the gap structure. In nematic phase, superfluid density is suppressed because of the reduction of DOS near the Fermi surface, resulting in a decrease of $T_\textrm{c}$. Although an origin of the disappearance of nematic order by Te substitution is difficult to infer from our results alone, a recent ARPES study on our films revealed an upward shift of the $d_{xy}$ orbital with increasing Te content because of the change of the chalcogen height.\cite{Nakayama2021} The approach of the $d_{xy}$ orbital to Fermi energy reduces the relative contribution of the $d_{xz/yz}$ orbitals which is important for nematicity, resulting in instability of nematic order. Also, we should comment on the difference in the behaviors of the $T_\textrm{c}$ and other quantities such as DOS between Fe(Se,Te) films and Fe(Se,S) films, which do not show a drastic increase of $T_\textrm{c}$ when nematic order disappears.\cite{Nabeshima2018a} In Fe(Se,S) films, our recent $\mu$-SR study showed the appearance of the short range magnetic order at high S contents,\cite{Nabeshima2018a,Nabeshima2021a} which is absent in Te substituted films. This may be one of the possible origins to explain the contrasting behavior of $T_\textrm{c}$ in Fe(Se,Te) and Fe(Se,S) films. Another important feature for S-substituted system is the weakening of the electronic correlation with increasing S content,\cite{Coldea2019,Kreisel2020a} which is again, in contrast to Te substituted films.\cite{Nakajima} In summary, phase diagrams of the FeSe$_{1-x}$Te$_x$ ($x=0-0.5$) films can be explained by considering their band structure as a primary factor. This indicates that band deformation in the nematic phase rather than the existence of the nematic order itself or possible nematic fluctuations developing near the quantum critical point is the predominant factor on superconductivity. Topological nature established in highly Te substituted materials might be another important factor,\cite{Zhang2018a} which is the subject of future works.



\section{Conclusion}
To conclude, we measured the complex conductivity of  FeSe$_{1-x}$Te$_x$ ($x=0-0.5$) films below $T_\textrm{c}$ combining coplanar waveguide resonator and cavity perturbation techniques. In the presence of nematic order, the temperature dependences of superfluid density and quasiparticle scattering time were qualitatively distinct from those of films without nematic order. This difference indicates that nematic order strongly influences the formation of nodes or gap minima in its superconducting gap structure. Conversely, the proportionality between $T_\textrm{c}$ and $\lambda_0^{-2}$ was observed irrespective of the presence or absence of nematic order, suggesting that the amount of superfluid exerts a more direct influence on the $T_\textrm{c}$ of Fe(Se,Te) than the nematic order itself. Combining those results with the band dispersion calculated based on DFT, we propose that the change of the Fermi surface in the nematic phase is the main factor for changes of $T_\textrm{c}$ and the corresponding superconducting gap structure in Fe(Se,Te).

\section*{Acknowledgments}
This work was supported by JSPS KAKENHI Grant Numbers  JP20H05164, JP19K14661.

\bibliographystyle{apsrev4-1-noarXiv}
%

\end{document}